\RequirePackage{lineno}
\documentclass[prl,superscriptaddress,twocolumn,showpacs,preprintnumbers,amsmath,amssymb]{revtex4}

\usepackage[pdftex]{graphicx}
\usepackage{dcolumn}
\usepackage{bm}

\usepackage{epsfig}
\usepackage{wrapfig}

\def\jpsi  {\mbox {$J/\psi\,$}}
\def\kstar {\mbox {$K^{*0}\,$}}
\def\kstarabstract {\mbox {$K^*(892)^0\,$}}
\def\kplus {\mbox {$K^+$}}

\def\ks    {\mbox {$K^0_s\,$}}
\def\kshort{\mbox {$K^0_s\,$}}
\def\pt    {\mbox {$p_T\,$}}

\def\bzero {\mbox {$B^0$}}

\def\bplus {\mbox {$B^+$}}

\def\lamB  {\mbox {$\Lambda^0_b$}}
\def\lam  {\mbox {$\Lambda^0$}}
\def\lxy   {\mbox {$L_{xy}$}}

\def\bite{\begin{itemize}}
\def\eite{\end{itemize}}
\def\bnum{\begin{enumerate}}
\def\enum{\end{enumerate}}

\def\bplusToPsiKPlus {\mbox{$\bplus \rightarrow \jpsi \kplus$}}

\def\bzeroToPsiKStar {\mbox{$\bzero \rightarrow \jpsi \kstar$}}
\def\bzeroToPsiKStarabstract {\mbox{$\bzero \rightarrow \jpsi \kstarabstract$}}

\def\bzeroToPsiKShort{\mbox{$\bzero \rightarrow \jpsi \ks$}}
\def\lambToPsiLam{\mbox{$\lamB \rightarrow \jpsi \Lambda^0$}}

\newcommand{\GeV}{\ensuremath{\mathrm{Ge\kern -0.1em V}}}
\newcommand{\MeV}{\ensuremath{\mathrm{Me\kern -0.1em V}}}
\newcommand{\keV}{\ensuremath{\mathrm{ke\kern -0.1em V}}}

\newcommand{\ppbar}{\rm{p\bar{p}}}
\begin{document}

\title { \boldmath
 Measurement of $b$ hadron lifetimes in exclusive decays \\
 containing a $\jpsi$ in $\ppbar$ collisions at $\sqrt{s}=1.96~{\rm TeV}$}
                                                                                              
\affiliation{Institute of Physics, Academia Sinica, Taipei, Taiwan 11529, Republic of China} 
\affiliation{Argonne National Laboratory, Argonne, Illinois 60439, USA} 
\affiliation{University of Athens, 157 71 Athens, Greece} 
\affiliation{Institut de Fisica d'Altes Energies, Universitat Autonoma de Barcelona, E-08193, Bellaterra (Barcelona), Spain} 
\affiliation{Baylor University, Waco, Texas 76798, USA} 
\affiliation{Istituto Nazionale di Fisica Nucleare Bologna, $^{bb}$University of Bologna, I-40127 Bologna, Italy} 
\affiliation{Brandeis University, Waltham, Massachusetts 02254, USA} 
\affiliation{University of California, Davis, Davis, California 95616, USA} 
\affiliation{University of California, Los Angeles, Los Angeles, California 90024, USA} 
\affiliation{Instituto de Fisica de Cantabria, CSIC-University of Cantabria, 39005 Santander, Spain} 
\affiliation{Carnegie Mellon University, Pittsburgh, Pennsylvania 15213, USA} 
\affiliation{Enrico Fermi Institute, University of Chicago, Chicago, Illinois 60637, USA}
\affiliation{Comenius University, 842 48 Bratislava, Slovakia; Institute of Experimental Physics, 040 01 Kosice, Slovakia} 
\affiliation{Joint Institute for Nuclear Research, RU-141980 Dubna, Russia} 
\affiliation{Duke University, Durham, North Carolina 27708, USA} 
\affiliation{Fermi National Accelerator Laboratory, Batavia, Illinois 60510, USA} 
\affiliation{University of Florida, Gainesville, Florida 32611, USA} 
\affiliation{Laboratori Nazionali di Frascati, Istituto Nazionale di Fisica Nucleare, I-00044 Frascati, Italy} 
\affiliation{University of Geneva, CH-1211 Geneva 4, Switzerland} 
\affiliation{Glasgow University, Glasgow G12 8QQ, United Kingdom} 
\affiliation{Harvard University, Cambridge, Massachusetts 02138, USA} 
\affiliation{Division of High Energy Physics, Department of Physics, University of Helsinki and Helsinki Institute of Physics, FIN-00014, Helsinki, Finland} 
\affiliation{University of Illinois, Urbana, Illinois 61801, USA} 
\affiliation{The Johns Hopkins University, Baltimore, Maryland 21218, USA} 
\affiliation{Institut f\"{u}r Experimentelle Kernphysik, Karlsruhe Institute of Technology, D-76131 Karlsruhe, Germany} 
\affiliation{Center for High Energy Physics: Kyungpook National University, Daegu 702-701, Korea; Seoul National University, Seoul 151-742, Korea; Sungkyunkwan University, Suwon 440-746, Korea; Korea Institute of Science and Technology Information, Daejeon 305-806, Korea; Chonnam National University, Gwangju 500-757, Korea; Chonbuk National University, Jeonju 561-756, Korea} 
\affiliation{Ernest Orlando Lawrence Berkeley National Laboratory, Berkeley, California 94720, USA} 
\affiliation{University of Liverpool, Liverpool L69 7ZE, United Kingdom} 
\affiliation{University College London, London WC1E 6BT, United Kingdom} 
\affiliation{Centro de Investigaciones Energeticas Medioambientales y Tecnologicas, E-28040 Madrid, Spain} 
\affiliation{Massachusetts Institute of Technology, Cambridge, Massachusetts 02139, USA} 
\affiliation{Institute of Particle Physics: McGill University, Montr\'{e}al, Qu\'{e}bec, Canada H3A~2T8; Simon Fraser University, Burnaby, British Columbia, Canada V5A~1S6; University of Toronto, Toronto, Ontario, Canada M5S~1A7; and TRIUMF, Vancouver, British Columbia, Canada V6T~2A3} 
\affiliation{University of Michigan, Ann Arbor, Michigan 48109, USA} 
\affiliation{Michigan State University, East Lansing, Michigan 48824, USA}
\affiliation{Institution for Theoretical and Experimental Physics, ITEP, Moscow 117259, Russia}
\affiliation{University of New Mexico, Albuquerque, New Mexico 87131, USA} 
\affiliation{Northwestern University, Evanston, Illinois 60208, USA} 
\affiliation{The Ohio State University, Columbus, Ohio 43210, USA} 
\affiliation{Okayama University, Okayama 700-8530, Japan} 
\affiliation{Osaka City University, Osaka 588, Japan} 
\affiliation{University of Oxford, Oxford OX1 3RH, United Kingdom} 
\affiliation{Istituto Nazionale di Fisica Nucleare, Sezione di Padova-Trento, $^{cc}$University of Padova, I-35131 Padova, Italy} 
\affiliation{LPNHE, Universite Pierre et Marie Curie/IN2P3-CNRS, UMR7585, Paris, F-75252 France} 
\affiliation{University of Pennsylvania, Philadelphia, Pennsylvania 19104, USA}
\affiliation{Istituto Nazionale di Fisica Nucleare Pisa, $^{dd}$University of Pisa, $^{ee}$University of Siena and $^{ff}$Scuola Normale Superiore, I-56127 Pisa, Italy} 
\affiliation{University of Pittsburgh, Pittsburgh, Pennsylvania 15260, USA} 
\affiliation{Purdue University, West Lafayette, Indiana 47907, USA} 
\affiliation{University of Rochester, Rochester, New York 14627, USA} 
\affiliation{The Rockefeller University, New York, New York 10065, USA} 
\affiliation{Istituto Nazionale di Fisica Nucleare, Sezione di Roma 1, $^{gg}$Sapienza Universit\`{a} di Roma, I-00185 Roma, Italy} 

\affiliation{Rutgers University, Piscataway, New Jersey 08855, USA} 
\affiliation{Texas A\&M University, College Station, Texas 77843, USA} 
\affiliation{Istituto Nazionale di Fisica Nucleare Trieste/Udine, I-34100 Trieste, $^{hh}$University of Trieste/Udine, I-33100 Udine, Italy} 
\affiliation{University of Tsukuba, Tsukuba, Ibaraki 305, Japan} 
\affiliation{Tufts University, Medford, Massachusetts 02155, USA} 
\affiliation{Waseda University, Tokyo 169, Japan} 
\affiliation{Wayne State University, Detroit, Michigan 48201, USA} 
\affiliation{University of Wisconsin, Madison, Wisconsin 53706, USA} 
\affiliation{Yale University, New Haven, Connecticut 06520, USA} 
\author{T.~Aaltonen}
\affiliation{Division of High Energy Physics, Department of Physics, University of Helsinki and Helsinki Institute of Physics, FIN-00014, Helsinki, Finland}
\author{B.~\'{A}lvarez~Gonz\'{a}lez$^v$}
\affiliation{Instituto de Fisica de Cantabria, CSIC-University of Cantabria, 39005 Santander, Spain}
\author{S.~Amerio}
\affiliation{Istituto Nazionale di Fisica Nucleare, Sezione di Padova-Trento, $^{cc}$University of Padova, I-35131 Padova, Italy} 

\author{D.~Amidei}
\affiliation{University of Michigan, Ann Arbor, Michigan 48109, USA}
\author{A.~Anastassov}
\affiliation{Northwestern University, Evanston, Illinois 60208, USA}
\author{A.~Annovi}
\affiliation{Laboratori Nazionali di Frascati, Istituto Nazionale di Fisica Nucleare, I-00044 Frascati, Italy}
\author{J.~Antos}
\affiliation{Comenius University, 842 48 Bratislava, Slovakia; Institute of Experimental Physics, 040 01 Kosice, Slovakia}
\author{G.~Apollinari}
\affiliation{Fermi National Accelerator Laboratory, Batavia, Illinois 60510, USA}
\author{J.A.~Appel}
\affiliation{Fermi National Accelerator Laboratory, Batavia, Illinois 60510, USA}
\author{A.~Apresyan}
\affiliation{Purdue University, West Lafayette, Indiana 47907, USA}
% ...
\collaboration{CDF Collaboration\footnote{With visitors from $^a$University of Massachusetts Amherst, Amherst, Massachusetts 01003,
$^b$Istituto Nazionale di Fisica Nucleare, Sezione di Cagliari, 09042 Monserrato (Cagliari), Italy,
$^c$University of California Irvine, Irvine, CA  92697, 
$^d$University of California Santa Barbara, Santa Barbara, CA 93106
$^e$University of California Santa Cruz, Santa Cruz, CA  95064,
$^f$CERN,CH-1211 Geneva, Switzerland,
$^g$Cornell University, Ithaca, NY  14853, 
$^h$University of Cyprus, Nicosia CY-1678, Cyprus, 
$^i$University College Dublin, Dublin 4, Ireland,
$^j$University of Fukui, Fukui City, Fukui Prefecture, Japan 910-0017,
$^k$Universidad Iberoamericana, Mexico D.F., Mexico,
$^l$Iowa State University, Ames, IA  50011,
$^m$University of Iowa, Iowa City, IA  52242,
$^n$Kinki University, Higashi-Osaka City, Japan 577-8502,
$^o$Kansas State University, Manhattan, KS 66506,
$^p$University of Manchester, Manchester M13 9PL, England,
$^q$Queen Mary, University of London, London, E1 4NS, England,
$^r$Muons, Inc., Batavia, IL 60510,
$^s$Nagasaki Institute of Applied Science, Nagasaki, Japan, 
$^t$National Research Nuclear University, Moscow, Russia,
$^u$University of Notre Dame, Notre Dame, IN 46556,
$^v$Universidad de Oviedo, E-33007 Oviedo, Spain, 
$^w$Texas Tech University, Lubbock, TX  79609, 
$^x$IFIC(CSIC-Universitat de Valencia), 56071 Valencia, Spain,
$^y$Universidad Tecnica Federico Santa Maria, 110v Valparaiso, Chile,
$^z$University of Virginia, Charlottesville, VA  22906,
$^{aa}$Yarmouk University, Irbid 211-63, Jordan,
$^{ab}$Universidad Aut\'onoma de Madrid, Cantoblanco, 28049, Madrid.
$^{ii}$On leave from J.~Stefan Institute, Ljubljana, Slovenia, 
}}
\noaffiliation

\date{\today}



\begin{abstract}
We report on a measurement of $b$-hadron lifetimes in the fully reconstructed decay modes 
\bplusToPsiKPlus, \bzeroToPsiKStarabstract, \bzeroToPsiKShort, and \lambToPsiLam\ using data corresponding to an integrated luminosity of 4.3 ${\rm fb}^{-1}$, collected by the CDF II detector at the Fermilab Tevatron. 
The measured lifetimes are
$\tau$(\bplus) = [$1.639 \pm 0.009 ~({\rm stat}) \pm 0.009~{\rm (syst)] ~ ps}$,
$\tau$(\bzero) = [$1.507 \pm 0.010 ~({\rm stat}) \pm 0.008~{\rm (syst)] ~ ps}$ and
$\tau$(\lamB) = [$1.537 \pm 0.045 ~({\rm stat}) \pm 0.014~{\rm (syst)] ~ ps}$. The lifetime ratios are
$\tau$(\bplus)/$\tau$(\bzero) = [$1.088  \pm 0.009~({\rm stat})\pm 0.004~({\rm syst})$] and
$\tau$(\lamB)/$\tau$(\bzero) = [$1.020   \pm 0.030~({\rm stat})\pm 0.008~({\rm syst})]$. 
These are the most precise determinations of these quantities from a single experiment.

\end{abstract}
\maketitle
\hspace{-.75 cm}





\newcommand{\imag}{\Im}

The lifetime of ground-state hadrons containing a $b$ quark and
lighter quarks is largely determined by the charged weak decay of the
$b$ quark. 
Interactions involving the lighter quarks, referred
to as spectator processes, alter $b$-hadron lifetimes 
at approximately the 10\% level. Lifetimes are important to probe
our understanding of the low-energy strong interaction. 
While precise predictions for $b$-hadron lifetimes are difficult to calculate, 
ratios are predicted with fairly high accuracy by the Heavy Quark Expansion 
(HQE)~\cite{NLOQCD}. 
This framework of theoretical calculation is used to predict low energy QCD effects in many 
flavor observables. For example, HQE predicts the decay-width of $B_s$ mesons to final states common to $B^0_s$ and $\bar{B^0_s}$, 
$\Gamma_{12}^s$, which enters
the decay-width difference in the $B^0_s$ system and several $CP$ 
violation effects. The measurement of lifetime ratios 
provides a simple and accurate way to test the HQE framework as non 
standard model effects are expected to be highly suppressed in lifetimes.
The ratio $\tau(B^+)/\tau(B^0)$, $R_+$, (charge conjugates are implied throughout) is 
predicted to be in the range 1.04-1.08~\cite{NLOQCD,PDG,Bigi,Subleading}. 
Predictions for the ratio $\tau(\Lambda^0_b)/\tau(B^0)$, $R_\Lambda$, in HQE, which do not presently 
incorporate next-to-leading order QCD corrections, lie in the range $0.88\pm0.05$~\cite{PDG,Subleading,LBOriginals}. 
The first measurements of the $\Lambda^0_b$ lifetime have been at the lower end 
of that range. However, recent high precision measurements by the CDF experiment~\cite{ref:Mark, ref:Petar} based on 1.0 fb$^{-1}$ of data, are significantly higher than previous results. 
It's therefore useful to keep pursuing lifetime
measurements with increased precision to settle the issue.
In this letter we report precise measurements of $b$-quark meson lifetimes using the 
channels \bplusToPsiKPlus, \bzeroToPsiKStar, and \bzeroToPsiKShort, in addition to the 
lifetime of the $\Lambda^0_b$ baryon using the \lambToPsiLam\ decay channel.  
Our data sample corresponds 
to an integrated luminosity of 4.3 fb$^{-1}$ and consists of $\ppbar$ collisions
at a center of mass energy $\sqrt{s}=1.96~{\rm TeV}$ collected by the CDF II detector at the Fermilab
Tevatron. The measurement reported here
improves the previous CDF measurement~\cite{ref:Mark} of the $\Lambda_b^0$ lifetime by 
updating it with significantly more data. In all decay modes, the decay 
position of the $b$ hadron is estimated using only $J/\psi$ decay products so 
that differences in decay time resolution between channels is reduced and certain 
systematic uncertainties cancel in ratios of lifetimes. 

The components of the CDF II detector relevant to this analysis are
described briefly here.
Charged particles are reconstructed using six layers of silicon microstrip detectors with radii between 2.4 cm and 23 cm \cite{SVXII} and an open-cell drift chamber called the central outer tracker
(COT) \cite{COT}. These are immersed in a
1.4~T solenoidal magnetic field and cover the range $|\eta| \leq 1$,
where $\eta$ is the pseudorapidity defined as $\eta \equiv - \ln \tan
(\theta/2)$, and $\theta$ is the polar angle~\cite{refsystem}. 
Four layers of planar drift chambers (CMU) \cite{CMU} detect muons with
$ p_T > 1.4~{\rm GeV/c} $ within $|\eta|<0.6$.
Additional chambers and scintillators (CMX) \cite{CMX} 
cover $0.6 < |\eta| < 1.0$ for muons with $ p_T > 2.0~{\rm GeV/c}$.

The reconstruction of $b$-hadron candidates begins with the collection of $J/\psi \rightarrow \mu^+ \mu^-$
candidates using a dimuon trigger. The extremely
fast tracker (XFT) \cite{XFT} uses COT hit information to measure the
transverse momentum and azimuthal direction of charged tracks. Events with $J/\psi \rightarrow \mu^+ \mu^-$ candidates are recorded for further analysis if two or more extrapolated tracks are matched to CMU or CMX track segments, opposite-charge and opening-angle requirements are met, and the $J/\psi$ candidate has mass in the range 2.7 to 4.0 GeV/$c^2$. After offline reconstruction, tracks corresponding to two triggered muon candidates are constrained
to originate from a common vertex to make a $J/\psi \rightarrow \mu^+ \mu^-$ candidate. To
ensure a high-quality vertex for the lifetime measurement, each muon
track is required to have at least three hits in the silicon 
system. The reconstructed $\mu^{+}\mu^{-}$ invariant mass is required
to be in the range $3.014 < m(\mu\mu) < 3.174$~GeV/c$^2$. The $b$ hadron is assumed to originate from the average beamspot
determined as a function of time using inclusive jet data.
The primary vertex for a given event is the $x-y$ position of this 
beamspot at the average $z$ coordinate of the muon tracks at their 
closest approach to the beamline. The typical beamline size is $\approx$ 30 $\mu m$ in $x-y$, and this dominates the uncertainty on the decay length. 
The projection of the transverse decay vector onto the $b$--hadron $p_T$ direction, $L_{xy}$, and its uncertainty, $\sigma_{xy}$, are also obtained and are used to estimate the proper decay time, $ct = \frac{M L_{xy}}{\pt}$, 
and its uncertainty $\sigma^{ct}$, where $M$ and $\pt$ are 
the mass and transverse momentum of the $b$ hadron. The uncertainties in the primary and $J/\psi$ vertices, and the transverse momentum are all included in $\sigma^{ct}$ which has a typical value around 0.1 ps. Uncertainties in transverse momentum have a
negligible effect on $ct$ measurement, in comparison to the uncertainty on the vertex positions.

 We reconstruct $K^{*0} \rightarrow K^+ \pi^-$, $K^{0}_s \rightarrow \pi^+ \pi^-$, and $\Lambda^0 \rightarrow p \pi^-$ candidates from pairs of
oppositely-charged tracks fit to a common vertex.
As $\kshort$ and $\lam$ decays can occur outside some layers of the silicon
system due to their long lifetime, their tracks are not required to
have silicon hits. 
The fitted mass is 
required to be in a mass window; for the $K^{*0}$ this window is 
$0.84 < m(K\pi) < 0.96$ GeV/c$^2$, (the lower range is selected in order to 
avoid reflections from the $\phi\rightarrow K^+K^-$, where one kaon is misreconstructed as a pion), 
for the $K^0_s$ it is
$0.473 < m(\pi\pi) <  0.523$ GeV/c$^2$, and for the $\Lambda^0$ it is $1.107 <m (p\pi) < 1.125$ GeV/c$^2$. This corresponds to approximately $\pm 3 \sigma$, where $\sigma$ is the mass resolution
of the reconstructed signal.
We suppress $K^0_s$ and $\Lambda^0$ cross
contamination by rejecting $K^0_s$ ($\Lambda^0$) candidates with
proton-pion (pion-pion) invariant mass consistent 
with $\Lambda^0$ ($K^0_s$).
We reconstruct the $b$-hadrons by performing a kinematic fit of all $b$-hadron final state tracks to the
appropriate topology: two spatially separated vertices in the case of  $\Lambda_b^0 \rightarrow J/\psi \Lambda^0$ and $B^0\rightarrow J/\psi K^0_s$,
one vertex in all other cases. A mass constraint is applied in the $J/\psi$ fit, 
and the reconstructed momenta of the $K^0_s$ and $\Lambda^0$ are required to point back to the $J/\psi$ vertex.
We exclude candidates with $\sigma^{ct} > 100~\mu m$ to ensure well measured vertices.
Additional selection requirements implying consistency with the
fit assumptions (common vertex or vertices, mass and pointing constraints) are also applied.
Further selection requirements on the transverse momenta of the $b$-hadrons and daughter particles, invariant mass of the \kshort, \kstar, and $\Lambda^0$, the vertex probability of the $b$-hadrons, and the \lxy\ significance of the \kshort and $\Lambda^0$ were obtained via an optimization
procedure, which maximizes the quantity $\mathcal{S}/\sqrt{\mathcal{S}+\mathcal{B}}$ over all of the selection requirements. 
The number 
of signal events ($\mathcal{S}$) is estimated from simulation and 
the number of background events ($\mathcal{B}$) from the mass sidebands in data. Sidebands are events away from the mass peak and form a sample of pure background.

\begin{figure*} [htb]
\centerline{\makebox{
    \includegraphics[width=0.35\textwidth]{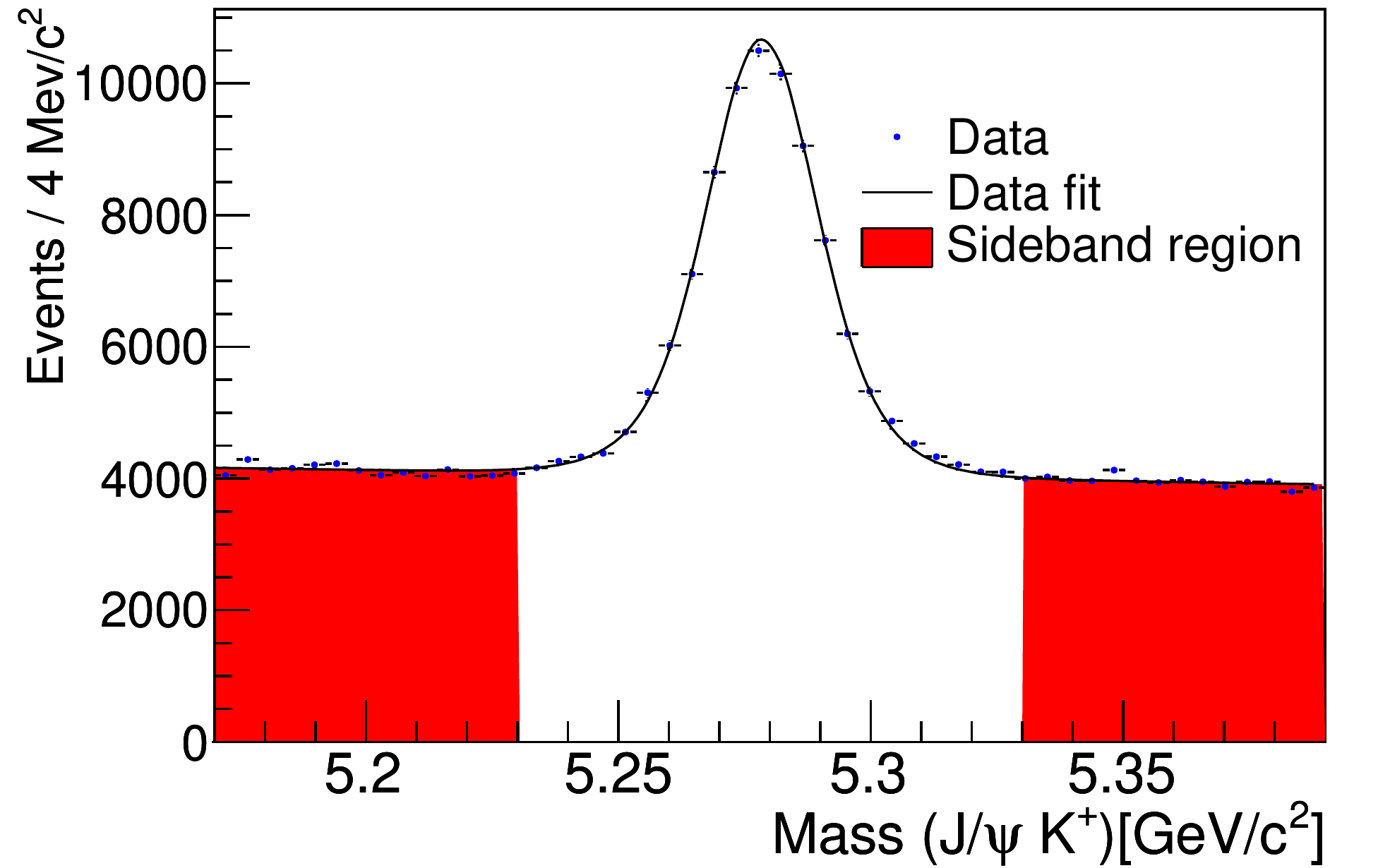}
   \unitlength=0.35\textwidth
   \put(-0.75,0.5){(a)}
    \includegraphics[width=0.35\textwidth]{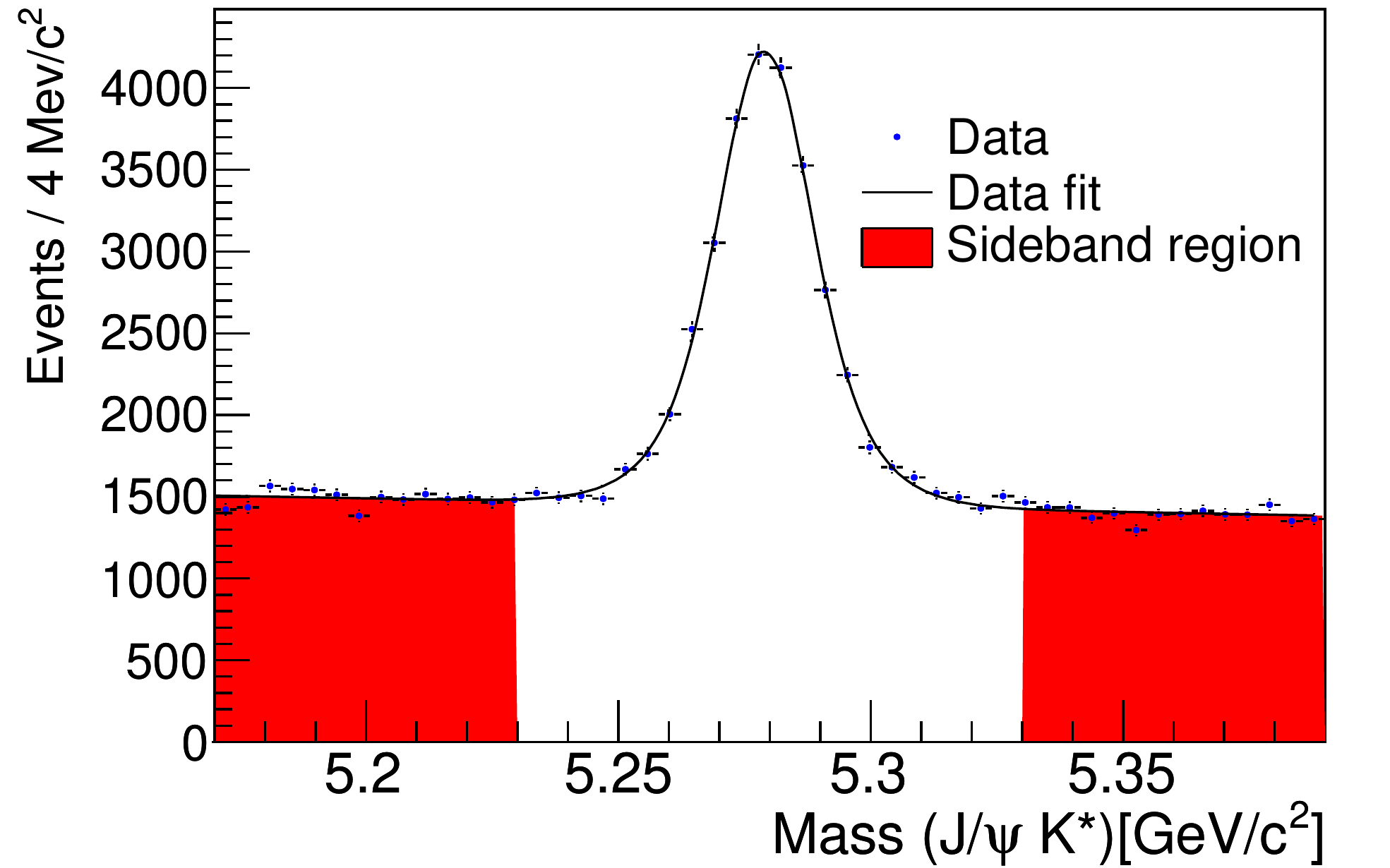}
   \unitlength=0.35\textwidth
   \put(-0.75,0.5){(b)}
    \includegraphics[width=0.35\textwidth]{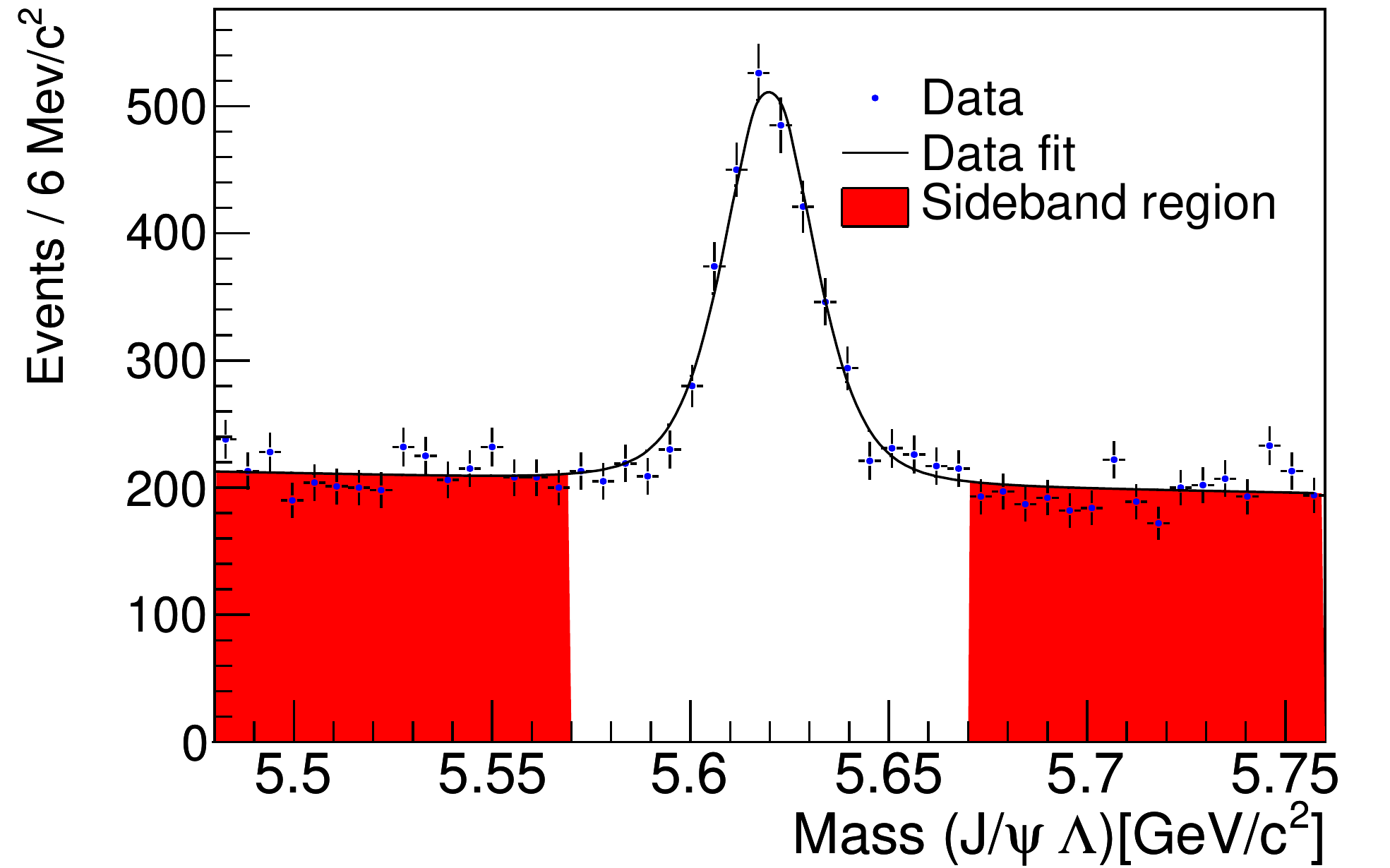}
   \unitlength=0.35\textwidth
   \put(-0.75,0.5){(c)}
}}
\caption{ Invariant mass together with mass fit projection for 
(a) $B^{+} \rightarrow J/\psi K^{+} $, (b) $B^0 \to J/\psi K^*$
, and (c) $ \Lambda^{0}_{b}\rightarrow J/\psi \Lambda^{0} $.}
\label{fig:mass}
\end{figure*}

For \bplus\ and \bzero\ modes, only 
candidates with a reconstructed $B$ mass between  5.17 and 5.39 GeV/c$^2$ 
are used for the lifetime measurements. 
For the \lamB\ mode, the mass range is set to 5.48 $-$ 5.76 GeV/c$^2$. These ranges provide a sufficient number of events in the sideband regions to constrain the background shape while avoiding regions where the mass distribution has complex structure. The invariant mass distributions for are shown in Fig.~\ref{fig:mass}, where the sideband regions are indicated.  The hadron masses are consistent with world average values. We observe the following yields of signal events: 45000 $\pm$ 230 (\bplus), 16860 $\pm$ 140 (\bzeroToPsiKStar), 12070 $\pm$ 120 (\bzeroToPsiKShort), and 1710 $\pm$ 50(\lamB).
\label{section:fittingModel}
The lifetimes are extracted using an unbinned maximum likelihood method. The
likelihood function ${\cal L}$ is multivariate, and is based on the probability of observing a candidate $i$ with reconstructed mass, $m_i$, decay time, $ct_i$, decay time uncertainty, $\sigma^{ct}_i$, and mass uncertainty, $\sigma^{m}_i$. The PDF is factorized in the following form:
\begin{eqnarray}
{\cal L} & = & \displaystyle\prod_i [ f_s \cdot  P_m^s(m_i|\sigma^{m}_i) \cdot T_t^s (ct_i|\sigma^{ct}_i) \cdot S_{\sigma^{ct}}^s (\sigma^{ct}_i) \\ \nonumber 
& + &  (1-f_s) \cdot P_m^b (m_i) \cdot T_t^b (ct_i|\sigma^{ct}_i) \cdot S_{\sigma^{ct}}^b (\sigma^{ct}_i) ],
\end{eqnarray}
where $P_m$, $T_{ct}$, and $S_{\sigma^{ct}}$ and are the normalized probability density functions (PDF) for observables $m_i$, $ct_i$ and $\sigma^{ct}_i$, the superscripts $s$ or $b$ refer to the PDF for signal or background candidates, respectively, and $f_s$ is the fraction of signal events. The PDF $S_{\sigma^{ct}}$ is substantially different for signal and 
background events and therefore needs to be taken into account as 
discussed in Ref.~\cite{punzi}.  A PDF term for $\sigma^{m}_i$ can be ignored as the distribution 
of $\sigma^m$ is observed to be similar for both signal and 
background and hence represents a constant in the log-likelihood. 

 The signal mass distribution, $P_m^s$, is modeled as
\begin{equation}
P_m^s = \sum_i \sum_{k=1}^2 \frac{F_K exp\left(-\frac{(m_0-m_i)^2}{2(u_k\sigma^m_i)^2}\right)}{\sqrt{2\pi}u_k\sigma^m_i} ,
\end{equation} where $F_1 + F_2 =1$, $m_0$ is the mass of the hadron, and $u_k$ are scale factors to account for the misestimation of the mass resolutions. We find that two Gaussians are sufficient to model the data. The background mass distribution, $P_m^b$, is modeled as
a linear function.

The signal $ct$ distribution, $T_{ct}^s$, is modeled 
by an exponential ($e^{- c t_i / c \tau} / c \tau$) convolved with a 
detailed detector $ct$-resolution model, $\cal{R}$. 
The background $ct$ distribution, $T_{ct}^b$, has four components: a $\delta-$function 
convolved with ${\cal R}$ to account for backgrounds from prompt $J/\psi$'s
originating from the primary vertex, and one 
increasing and two decreasing exponentials that account for mismeasured decay vertices 
and background from other heavy-flavor decays. These exponential components are convolved with a single Gaussian of 
width $\sigma^{ct}_i$ multiplied by a scale factor. 
The relative contribution of each background component is determined by 
the data. The parameters of the background model are mainly determined 
from the candidates in the mass sidebands. Studies of inclusive $b$--hadron 
decays have shown that after the selection requirements the 
contamination from other $b$ decays is very low and, furthermore, that 
the mass distribution of the long lived background components 
is flat in the fitted mass range, and hence the mass sidebands can 
provide a realistic background model for candidates in the signal mass range.

The same resolution model, ${\cal R}$, 
is used for signal and prompt background events. The detector 
resolution is based upon a Gaussian with width $s_j\cdot\sigma^{ct}_i$, where $s_j$ is the scale factor that accounts for the misestimation the $\sigma^{ct}_i$.
Motivated by a study of resolution in an inclusive sample 
of $J/\psi$ events, where prompt $J/\psi$ events dominate,
$\cal R$  is modeled as ${\cal R}= \sum_{j=1}^3 {f_j}/\left({\sqrt{2\pi}s_j\sigma^{ct}_i}\right)\cdot\exp\left({-{t^2}/{2(s_j \sigma^{ct}_i)^2}}\right)$, where $f_1+f_2+f_3 = 1$. The Gaussians are centered at zero as no evidence of an offset in the data samples is observed.
Small differences in ${\cal R}$ arise
between decay channels due to different 
$\chi^2$ distributions for the vertex fits 
of decays with different number of tracks. 
Therefore the parameters $f_j$ and $s_j$ are 
obtained separately for each channel from a fit to 
data in the mass sidebands. 
This yields an accurate determination 
of $\cal R$ since the background events
are primarily expected to originate from the interaction vertex.

\begin{figure*} [htb]
\centerline{\makebox{
    \includegraphics[width=0.35\textwidth]{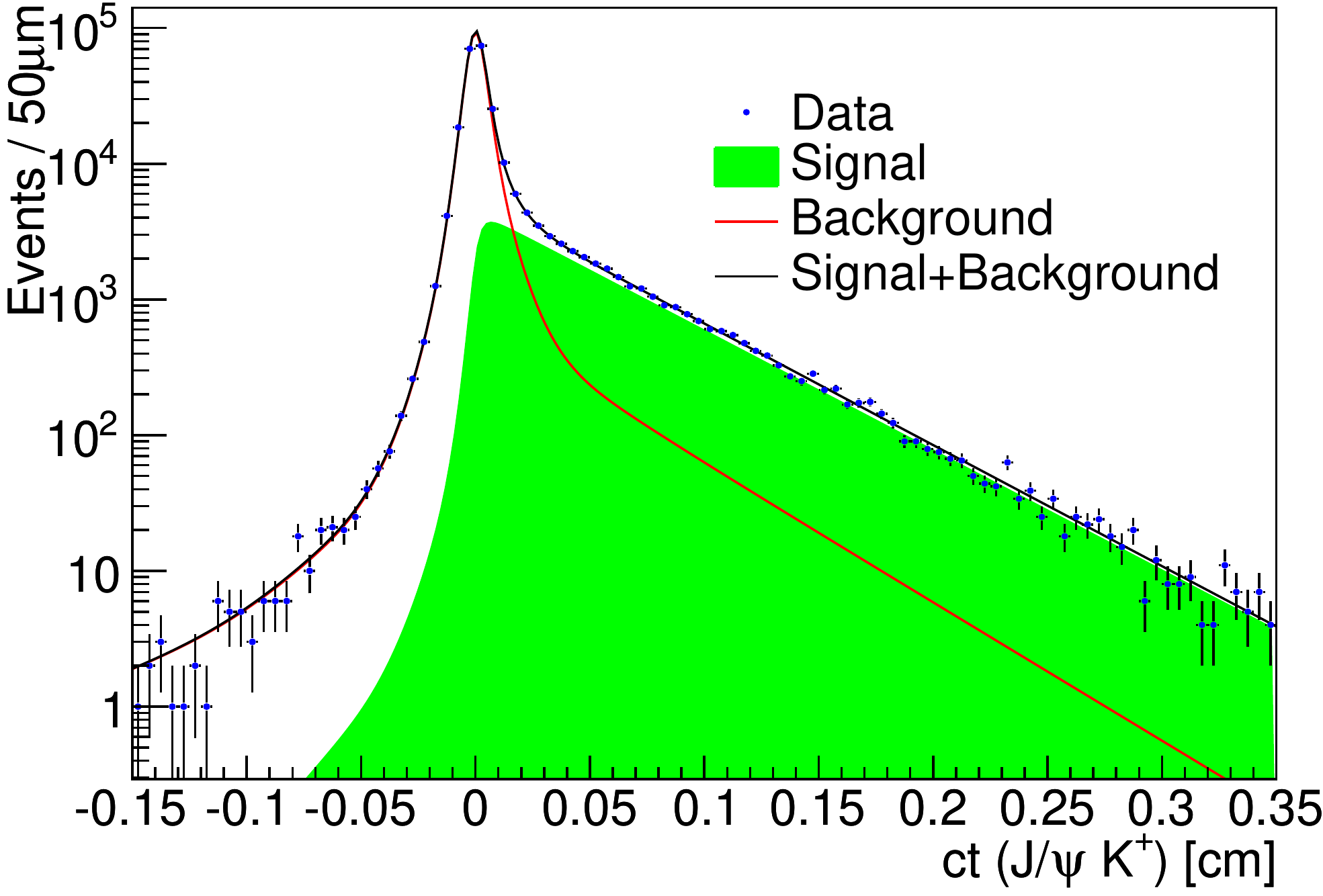}
  \unitlength=0.35\textwidth
    \put(-0.8,0.5){(a)}
    \includegraphics[width=0.35\textwidth]{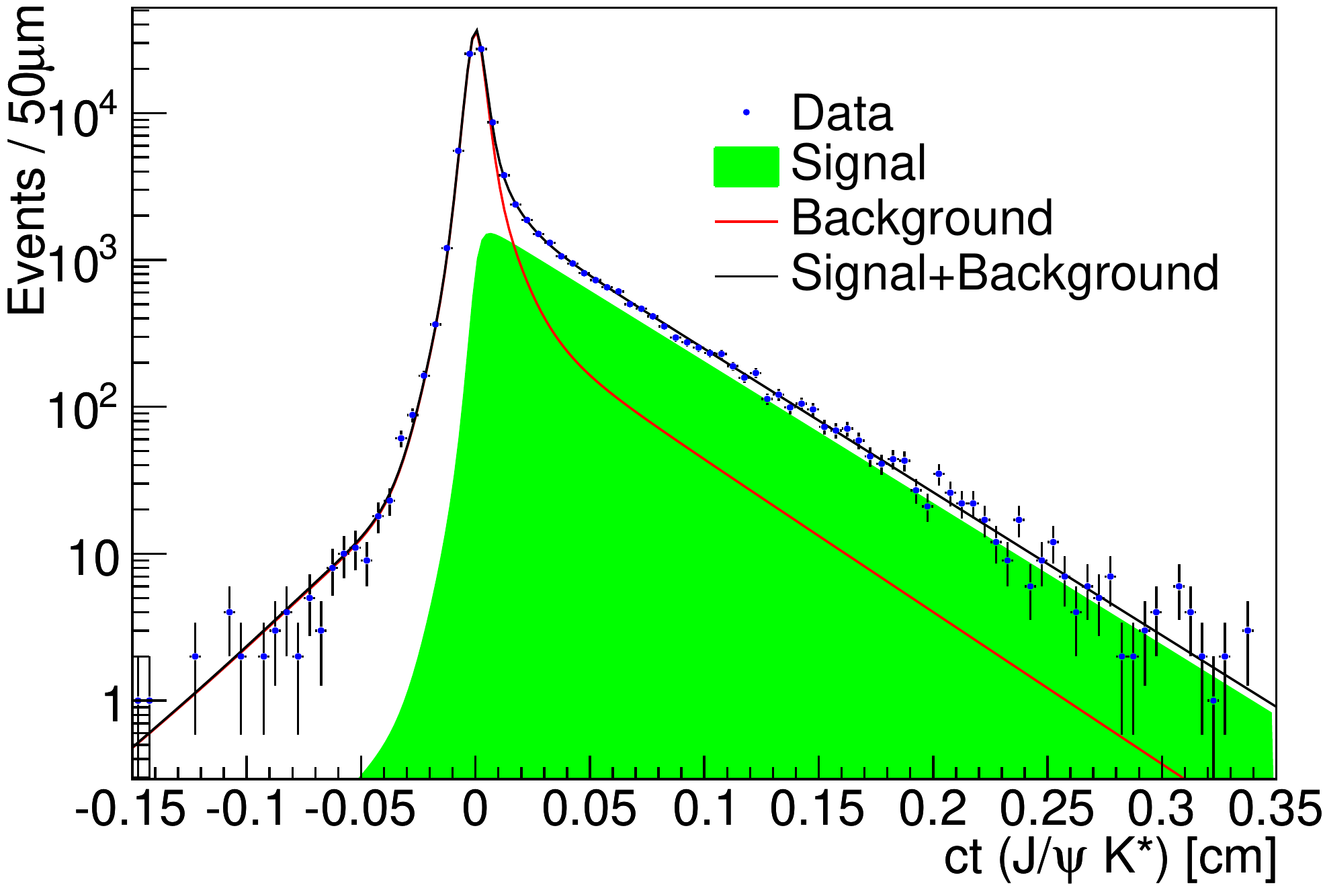}
    \unitlength=0.35\textwidth
    \put(-0.8,0.5){(b)}
    \includegraphics[width=0.35\textwidth]{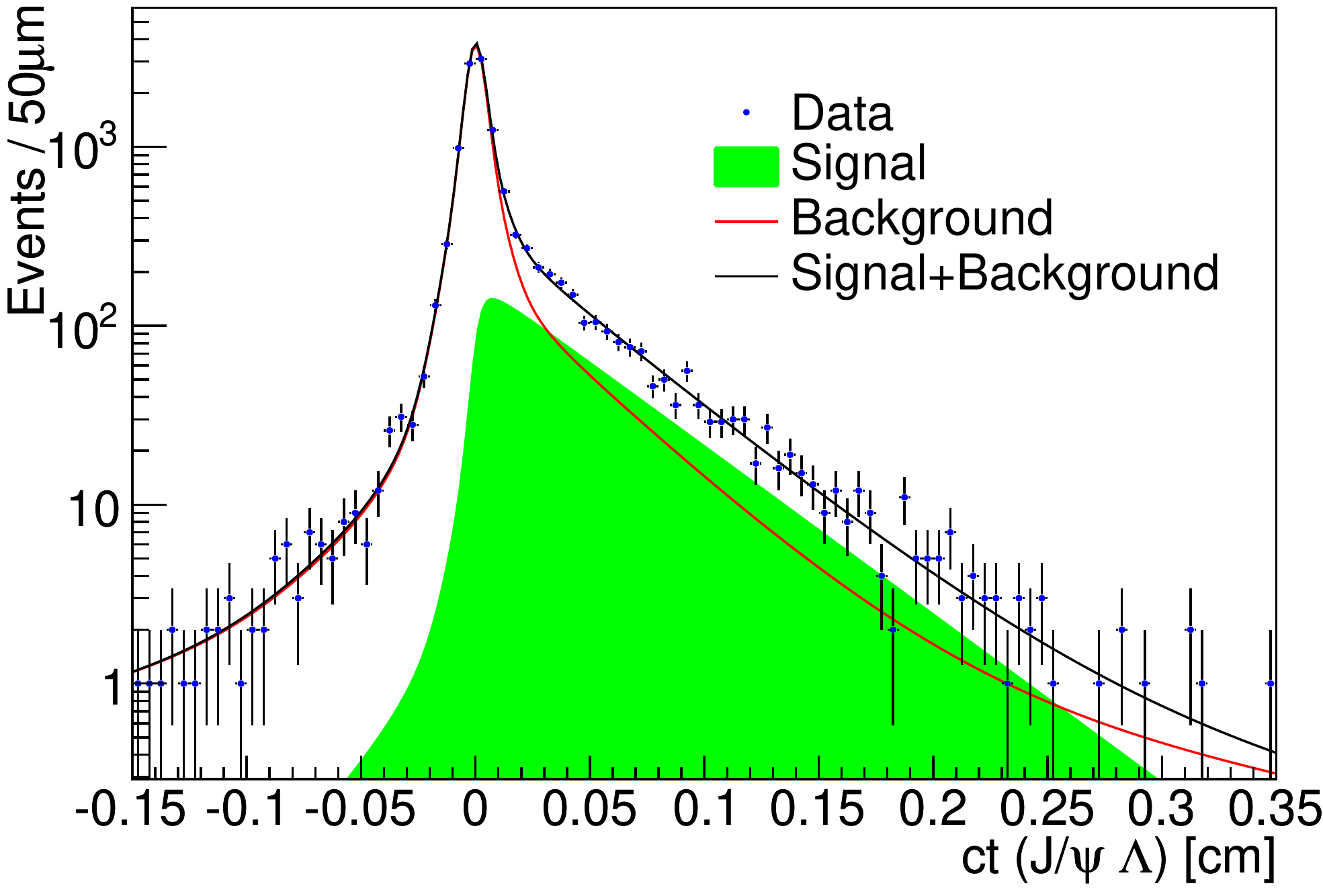}
    \unitlength=0.35\textwidth
    \put(-0.8,0.5){(c)}
}}
\caption{ Decay time distributions for (a) $B^{+} \rightarrow J/\psi K^{+}$, (b) $B^0 \to J/\psi K^{*}$, and (c) $\Lambda^0_b\rightarrow J/\psi \Lambda^0$ candidates.
}
\label{fig:Bpfit3c}
\end{figure*}

The functional form of the PDF $S_{\sigma^{ct}}$ is determined 
empirically using data in the B hadron mass sidebands as indicated on Fig.~\ref{fig:mass}. The parameters of the function, which are different for 
signal and background are determined from the final fit to data. After the resolution model parameters are determined from the mass-sideband 
only fit, the likelihood is calculated for each candidate and 
the product is maximized in each of the four channels to extract
the lifetime, signal yield and the other 
parameters required to describe the mass, background decay time, and $\sigma^{ct}$ distributions. Decay time projections of the likelihood function 
are compared with the data in Fig.~\ref{fig:Bpfit3c}.

We considered correlated and uncorrelated systematic uncertainties.
Correlated uncertainties affect all measured lifetimes identically, and cancel in ratios. We estimate uncertainties due
to any residual misalignments of the silicon detector using Monte
Carlo samples generated with radial displacements of individual
sensors (internal alignment) and relative translation and rotation of
the silicon detector with respect to the COT (global alignment). 
The XFT triggers on tracks assuming they originate from the center of the beam, which may introduce a bias for triggering long-lived decays. No indication of any bias was found in a study of the XFT response in a large sample of simulated events but a small uncertainty 
is assigned due to the limited statistical precision of the evaluation
method. 
The systematic uncertainty that results from ignoring the correlation
between reconstructed mass and $\sigma^{ct}$ in the likelihood is found to be
negligible.
The remainder of systematic uncertainties are treated as uncorrelated.
They were determined using 
pseudo-experiments
in which many statistical trials are generated 
according to alternate PDFs where the alternate parameters are derived from data. The shift in data due to the alternate PDFs were consistent with the shift observed with the pseudo-experiments.
As the time-resolution is determined from the prompt events, and the 
shape of those events is sensitive to the modeling of long-lived 
(positive and negative) background, uncertainties in the background 
modeling can affect the lifetime through the resolution function. We account 
for that uncertainty by including  an extra long-lived component in the 
background model. This alternate description produces a substantial change
in the  fraction of prompt events (approximately 7\%), and has a small 
but non-negligible effect on the lifetime. A further small uncertainty arising from the functional form of $\cal{R}$ is also assessed and included in the total resolution uncertainty.
To evaluate uncertainties in the mass model, alternate 
parametrizations, including a 2nd order polynomial for background, and a single Gaussian to describe signal events, were considered. Alternatives to the background PDF included extra long lived and Gaussian components. We determined the uncertainty due to the $\sigma^{ct}$ parametrization by using a reasonable alternate model. 
We also considered the effect of ignoring any differences between 
signal and background mass uncertainties by using distributions determined
from data to generate the values of the mass uncertainty in the pseudo-experiments.
We also determined the systematic uncertainty due to the presence of the Cabibbo suppressed 
channel $B^+ \to J/\psi \pi^+$ in the charged $B$ decays, and the effect 
of swapping the kaon and pion hypotheses in $K^{*0}$ reconstruction.
The possibility of a systematic biases caused by the $\sigma^{ct}$ and $p_T$ selection
requirements were found to be negligible.
The systematic uncertainties are summarized in Table
\ref{tab:syssum}.

{ \tiny

\begin{table*}[htbp]
\caption{Summary of systematic uncertainties. 
\label{tab:syssum}}

\hspace{-1.5cm}
\begin{tabular}{l c c c c c c }
\hline
\hline
 & $\jpsi K^+$ (fs) & $\jpsi\kstar$ (fs)   & $\jpsi\ks$ (fs)&   $\jpsi\lam$ (fs)  & ${R_+}$ & $R_{\Lambda}$ \\ \hline
Resolution function       &  2.5       &   3.5       &  3.0    &   8.9     & 0.0024  &  0.0061  \\
Background $ct$ model           &  1.0       &   2.3       &  4.1  &  4.6       & 0.0017  &  0.0034  \\
Mass model    &  2.8       &    2.8      &  2.8    & 2.8       & 0.0020  &  0.0017  \\
Proper decay time uncertainty             &  1.7       &    1.7      &     1.7 &  4.3       & 0.0010  &  0.0029  \\
Mass uncertainty                                  &  3.0       &   3.0       &     3.0 & 3.0       & 0.0020  &  0.0012  \\
\hline
Total uncorrelated               &    $\pm$ 5.2  & $\pm$ 6.2     &$\pm$ 6.8  & $\pm$ 11.7     &  0.0042 &    0.0079         \\

\hline
Alignment                                   & 6.7         &    6.7       &  6.7     &  6.7       & $-$  & $-$            \\
Cabibbo suppressed mode in \bplus           &  0.7       & $-$                 & $-$             & $-$               & 0.0004  &  $-$       \\
Swapped track assignment in $B^0$              & $-$                & 0.7        & $-$              & $-$               & $-$       &  $-$       \\
Possible trigger bias                       & 1.7        & 1.7         & 1.7     & 1.7       &$-$       & $-$        \\
$\sigma^{ct}$ -$m$  correlation          & 0.7        & 0.7         & 0.7     & 0.7       & $-$       &  $-$        \\
\hline
Total                                     &    $\pm$8.7  & $\pm$9.3     &$\pm$9.7  & $\pm$ 13.7     &  0.0043 &    0.0079         \\

\hline
\hline
\end{tabular}
\end{table*}

}

We measure $\tau(B^+)=[1.639\pm0.009\mathrm{(stat)}\pm0.009\mathrm{(syst)}]{\rm ~ps}$ and  
$\tau(B^0)=[1.507\pm0.010 \mathrm{(stat)}\pm0.008 \mathrm{(syst)}]{\rm ~ps}$ where the two 
$\bzero$ measurements have been combined. 
These results are consistent and of similar precision to the leading 
measurements from Belle~\cite{ref:BelleResult} 
which are $\tau(B^+)=[1.635 \pm 0.011\mathrm{(stat)} \pm 0.011\mathrm{(syst})$] ps and $\tau(B^0)=[1.534 \pm 0.008\mathrm{(stat)} \pm 0.010\mathrm{(syst)}$] ps. 
The similarities between the decay channels allow for the accurate 
determination of the ratio 
$R_+$ = [$1.088  \pm 0.009~({\rm stat})\pm 0.004~({\rm syst})$] 
which favors a slightly higher value than the current average of 1.071$\pm$0.009~\cite{PDG}. 
These results are consistent with the current HQE predictions, giving further 
confidence in this theoretical framework, and also provide an accurate test 
for future lattice QCD calculations.
For the $\Lambda^0_b$ we measure 
$\tau(\Lambda^0_b)=[1.537\pm ({\rm stat})0.045\pm({\rm syst}) 0.014]~{\rm ps}$ and 
$R_{\Lambda}=[1.020\pm 0.030\mathrm{(stat)} \pm0.008\mathrm{(syst)}$]. 
This measurement is the most precise measurement of 
$\tau(\Lambda^0_b)$ and is consistent with the previous 
CDF measurement in this decay channel of 
$\tau(\Lambda^0_b)= [1.593^{+0.083}_{-0.078}\mathrm{(stat)}\pm0.033\mathrm{(syst)}]~{\rm ps}$~\cite{ref:Mark}
but is more than 2$\sigma$ larger than the world average of 
$1.383^{+0.049}_{-0.048}$ ps and the previous 
CDF measurement ~\cite{ref:Petar}, performed 
on a different decay channel ($\Lambda_c^{\pm} \pi^{\mp}$): [$1.401\pm 0.046\mathrm{(stat)} \pm 0.035 \mathrm{(syst)}]~{\rm ps}$. 
The ratio is also higher 
than the predicted values of 0.88$\pm$0.95.
~In summary, we report the most precise determination of 
$\tau(B^+)/\tau(B^0)$. It is consistent with other 
measurements and the predicted value which gives 
confidence in the HQE framework for flavor observables. 
We also report the most precise measurement of $\tau(\Lambda^0_b)$, 
which supports a higher value than the world average and theory predictions.

\begin{acknowledgments}

We thank the Fermilab staff and the technical staffs of the participating institutions for their vital contributions. This work was supported by the U.S. Department of Energy and National Science Foundation; the Italian Istituto Nazionale di Fisica Nucleare; the Ministry of Education, Culture, Sports, Science and Technology of Japan; the Natural Sciences and Engineering Research Council of Canada; the National Science Council of the Republic of China; the Swiss National Science Foundation; the A.P. Sloan Foundation; the Bundesministerium f\"ur Bildung und Forschung, Germany; the World Class University Program, the National Research Foundation of Korea; the Science and Technology Facilities Council and the Royal Society, UK; the Institut National de Physique Nucleaire et Physique des Particules/CNRS; the Russian Foundation for Basic Research; the Ministerio de Ciencia e Innovaci\'{o}n, and Programa Consolider-Ingenio 2010, Spain; the Slovak R\&D Agency; and the Academy of Finland. 
\end{acknowledgments}


\begin{thebibliography}{99}
\bibitem{refsystem} CDF II uses a cylindrical coordinate system in which
$\phi$ is the azimuthal angle, $r$ is the radius from the nominal beamline,
and $z$ points in the proton beam direction with the origin at the center of
the detector. The transverse plane is the plane perpendicular to the $z$ axis.
The ${\hat x}$ axis points outward from the accelerator ring, and the ${\hat y}$ axis points upwards.
\end{thebibliography}
\end{document}